\documentclass{article}

\usepackage{amsmath}
\usepackage{latexsym}

\def\be{\begin{equation}}
\def\ee{\end{equation}}
\def\bea{\begin{eqnarray}}
\def\eea{\end{eqnarray}}
\def\({\left(}
\def\){\right)}
\def\<{\left<}
\def\>{\right>}

\def\[{\left[}
\def\]{\right]}

\begin{document}

\pagestyle{empty}
\vskip-10pt
\vskip-10pt
\hfill {\tt hep-th/0506036}
\begin{center}
\vskip 3truecm
{\Large\bf
Energy radiated from a fluctuating selfdual string
}\\ 
\vskip 2truecm
{\large\bf
Andreas Gustavsson\footnote{a.r.gustavsson@swipnet.se}
}\\
\vskip 1truecm
{\it Institute of Theoretical  Physics,
Chalmers University of Technology, \\
S-412 96 G\"{o}teborg, Sweden}\\
\end{center}
\vskip 2truecm
\noindent{\bf Abstract:}
We compute the energy that is radiated from a fluctuating selfdual string in the large $N$ limit of $A_{N-1}$ theory using the AdS-CFT correspondence. We find that the radiated energy is given by a non-local expression integrated over the string world-sheet. We also make the corresponding computation for a charged string in six-dimensional classical electrodynamics, thereby generalizing the Larmor formula for the radiated energy from an accelerated point particle. 
\vfill \vskip4pt

\eject
\newpage
\pagestyle{plain}
\section{Introduction}
M-theory on $(AdS_7\times S^4)_N$ is conjectured to correspond to the six-dimensional $A_{N-1}$, $(2,0)$-theory which has a superconformal symmetry times $SO(5)_R$, corresponding to the supersymmetries preserved by $AdS_7\times S^4$ \cite{Maldacena1}, \cite{Witten}. The conjecture states that we should be able to use classical supergravity if the characteristic length scale $\sim l_P N^{1/3}$ of $(AdS_7\times S^4)_N$ is much larger than the Planck length $l_P$, that is, if $N>>1$. At the conformal point we have $N$ parallel five branes on top of each other. If we separate one of the branes from the stack, we may have a membrane streching between the branes. The boundary of the membrane appears as a selfdual string in the boundary theory. For well separated five branes, this theory is described by $N$ tensormultiplets, interacting with $\frac{N(N-1)}{2}$ different types of tensile strings \cite{Strominger}. For well-separated five-branes we can think of the tensile degrees of freedom as strings whereas the massless degrees of freedom are tensormultiplet particles. 

The time-like Wilson surface observable, associated with a tensile string world-sheet, corresponds in the large $N$ limit to the area of the extremal membrane \cite{Maldacena2}. Accordingly fluctuations of the string should give rise to nonlinear waves in the membrane. These waves should in the large $N$ limit be determined by the condition that the area of the membrane is extremized, under the condition that the membrane is attached to the selfdual string. On the gravity side the energy radiated from the string goes out in the membrane. On the gauge theory side on the other hand, it goes out into the tensor fields living on the five brane. One should now recall that the AdS-CFT correspondence is holographic in the sense that the boundary theory is conjectured to be equivalent to M-theory on $AdS$ space, not to be supplemented by M-theory on $AdS$. It is thus wrong to think that the energy would go out into both the membrane and into the six-dimensional boundary theory, at least in the large $N$ limit. Each degree of freedom in the boundary theory should be possible to map to a corresponding degree of freedom in the theory in $AdS$. Having the correspondence between the Wilson surface observable and the extremal membrane, it is natural to think that the total energy radiated from the string corresponds to the increase of energy in this extremal membrane. An interesting question to ask is of course what would the differential cross section correspond to on the {\sl{AdS}} side? But we will not make any attempt to answer that question in this paper.

In the reference \cite{Mikhailov} a computation of the energy and momentum radiated from a `quark' in strongly coupled (large 't Hooft coupling) ${\cal{N}}=4$ superYang-Mills was made in the large $N$ limit where the string coupling constant is small. What was remarkable there was that the radiated energy-momentum is given by a local expression integrated over the world-line of the quark. By Lorentz invariance and dimensional analysis, any local expression for the radiation has to be proportional to the Lienard-Wiechert formula in linear classical electrodynamics. This was also obtained explicitly in that paper using a nonlinear wave solution. Here we will see that a corresponding agreement does not quite show up for the radiation from a selfdual string.

\section{The linearized wave in AdS space}
According to the AdS-CFT correspondence, a $k$-dimensional submanifold observable in the boundary theory, corresponds to a $k$-brane whose boundary is the $k$-submanifold and that extends into the AdS space. In the large $N$ limit the $k$-brane obeys the classical equation of motion which means that its area is extremized. We can cover half of the AdS space with Poincare coordinates $(y,x^{\mu})$, $y>0$. In these coordinates the metric is
\bea
ds^2 = \frac{1}{y^2}\(dy^2+\eta_{\mu\nu}dx^{\mu}dx^{\nu}\)
\eea
where $\eta_{\mu\nu} = $diag$(-1, 1, 1, ..., 1)$ is the Minkowski metric in the boundary theory. We will use Monge gauge for the $k$-brane, $x^A=(y,x^{\alpha})\equiv (y,t,x^a)\mapsto (Y,X^{\mu})=(y,t,x^a,X^i(y,t,x))$, $a=1,...,k-1$. In this gauge, the induced metric on the $k$-brane is
\bea
ds^2 = \frac{1}{y^2}\bigg(\(1+(\partial_yX)^2\)dy^2+2\partial_{\alpha}X.\partial_yXdx^{\alpha}dy+\(\eta_{\alpha\beta}+\partial_{\alpha}X.\partial_{\beta}X\)dx^{\alpha}dx^{\beta}\bigg)
\eea
To linear order, the equation of motion that we derive from the area functional for the $k$-brane,
\bea
A[X]=\int dy dt d^{k-1}x \sqrt{-g}
\eea
($g$ denotes the determinant of the induced metric) is
\bea
-\partial^2_{\alpha}{X}-\partial^2_y{X}+(k+1){y}^{-1}\partial_y{X}=0
\eea
We impose a boundary condition 
\bea
X(0,x)=\xi(x)
\eea
The solution may be expressed as
\bea
X(y,x)=\int d^kx' K(y,x;x')\xi(x')\label{Green}
\eea
where the Green's function (which really is a distribution) is given by
\bea
K(y,x;x')=c\frac{y^{k+2}}{\(-(x-x')^2-y^2\)^{k+1}}\label{Greenk}
\eea
To get an outgoing wave solution we should integrate over times $t'\leq t-\sqrt{y^2+\sum_{a=1}^{k-1}(x^a-x'^a)^2}$.

We can make the substitution $\zeta^{\alpha}=(x-x')^{\alpha}/y$ and rewrite the solution as 
\bea
X(y,x)=c\int d^k\zeta \frac{\xi(x-\zeta y)}{(-\zeta^2-1)^{k+1}}
\eea
In this form it is easy to find an expression for the normalization constant $c$ from the boundary condition, though all these integrals diverges, so to get a finite result we must really take $c=0$ to get $0\cdot\infty$ which could be anything. Fortunately we can determine exactly how much the integrals diverges, which enable us to remove the divergence in a mathematically precise way. The solution as we have presented it in terms of a Green's function does not exist in any dimension $k=0,1,2,...$. But if we in the Green's function Eq. (\ref{Greenk}) replace $k$ by any complex number $\lambda$ such that Re $\lambda<-1$, then that corresponding convolution integral with $\xi$ will exist. Furthermore that integral will be analytic in $\lambda$. We can thus continue the integral analytically almost all the way to the integer dimensions $\lambda=k$, where we will have poles. These poles may be removed by dividing by another analytic function which also has poles at those integer dimensions $k$. The mathematical theory for this can be found in Ref. \cite{Gelfand}, but we will review it here.

Define $x_+=x$ if $x>0$ and $x_+=0$ if $x\leq 0$. If $\lambda\in {\bf{C}}$ and Re $\lambda>0$ then $x_+^{\lambda-1}$ defines a distribution,
\bea
<x_+^{\lambda-1},\phi>=\int_0^{\infty} dx x^{\lambda-1}\phi(x)\label{one}
\eea
where $\phi$ is a test function (a smooth function with compact support). In particular then, for $\lambda=1$, we find that $x_+^0=\theta$ is the Heaviside step function. The distribution (\ref{one}) can be extended to any $\lambda\in {\bf{C}}$. For instance we have the relation
\bea
\int_0^{\infty} dx x^{\lambda-1}\phi(x)=-\frac{1}{\lambda}\int_0^{\infty} dxx^{\lambda}\partial\phi(x)
\eea
for Re $\lambda>0$ by doing an integration by parts. Furthermore both sides are analytic in $\lambda$. But the right-hand side is well-defined even for Re $\lambda+1>0$, even where the left-hand side is a divergent integral. It is now clear that the analytic continuation of the distribution $x_+^{\lambda-1}$ should be defined by such integrals that we obtain by doing integrations by parts. We then get poles at $\lambda=-k=0,-1,-2,...$ with residues $(-1)^k/k!\partial^k\delta$. If we then divide by $\Gamma(\lambda)$ which also has poles at $\lambda=-k=0,-1,-2,...$, with residues $(-1)^k/k!$, we obtain a distribution\footnote{These properties follow directly from the following properties of the gamma function 
\bea
\Gamma(1)&=&1\cr
\Gamma(\lambda+1)&=&\lambda\Gamma(\lambda)\cr
\Gamma(\lambda)\Gamma(1-\lambda)&=&\frac{\pi}{\sin(\pi \lambda)}
\eea} 
\bea
\partial^{\lambda}\delta=\frac{x_+^{-\lambda-1}}{\Gamma(-\lambda)} {\mbox{ ,  $\lambda\notin\{0,1,2,...\}$}} 
\eea
which will be analytic in $\lambda$ everywhere in the complex plane if we define $\partial^{\lambda}=\partial^k$ to be the usual derivative for $\lambda=k=0,1,2,...$, thus justifying the notation. For non-integer $\lambda$'s, this distribution defines a `fractional derivate'. We should note that although the ordinary derivative is local, the fractional derivative is not. It can be shown that the fractional derivative obeys $\partial^{\lambda}\partial^{\lambda'}=\partial^{\lambda+\lambda'}$. We also have the Leibniz rule,
\bea
\partial^{p}(fg)=\sum_{k=0}^{\infty}\({}^p_k\)\partial^{p-k} f \partial^k g
\eea
where we define the binomial coefficients for any complex number $p$ as
\bea
\({}^p_k\)\equiv\frac{p}{1}...\frac{p-k+1}{k}
\eea
To see this we just Taylor expand $g(y)$ around $x$ in the definition
\bea
\partial^p(f(x)g(x))\equiv\frac{1}{\Gamma(-p)}\int_x^{\infty}dy\frac{f(y)g(y)}{(x-y)^{p+1}}
\eea
We may interchange the roles of $f$ and $g$ and get a different looking expansion.

It is instructive to consider the wave solution for the case $k=1$. Then we have
\bea
X(y,t)=c\int_{1}^{\infty} d\zeta \frac{\xi(t-\zeta y)}{(\zeta^2-1)^{2}}
\eea
We have a singularity at $\zeta=1$ of degree $2$. Interpreting $c(\zeta-1)_+^{-2}$ as the distribution defined above with "$c=-4/\Gamma(-1)$" (which is $c=1/\infty=0$), we get
\bea
X(y,t)=\left.-4\frac{\partial}{\partial\zeta} \(\frac{\xi(t-\zeta y)}{(\zeta+1)^{2}}\)\right|_{\zeta=1}
\eea
which yields
\bea
X(y,t)=\xi(t-y)+y\xi'(t-y)
\eea
which is the linearized wave solution that was found in \cite{Mikhailov}. 

In dimension $k$, the linearized wave solution is
\bea
X^i(y,t,{\vec{x}})=c\int  d^{k-1}\vec{\zeta}\int_{\sqrt{|\vec{\xi}|^2+1}}^{\infty} d\zeta \frac{\xi(t-\zeta y,{\vec{x}}-{\vec\zeta}y)}{(\zeta^2-|{\vec{\zeta}}|^2-1)^{k+1}}
\eea
If we now make the simplifying assumption that $\xi(t,x)=\xi(t)$, we can very easily perform the integrals over $\vec{\zeta}$ to get
\bea
X^i(y,t)=c\int_{1}^{\infty} d\zeta \frac{\xi(t-\zeta y)}{\(\zeta^2-1\)^{\frac{k+3}{2}}}
\eea
Here we have a singularity at $\zeta=1$ of order $\frac{k+3}{2}$. Thus
\bea
X^i(y,t)=\left.(-2)^{\frac{k+3}{2}}c_{\frac{k+1}{2}}\(\frac{\partial}{\partial\zeta}\)^{\frac{k+1}{2}} \(\frac{\xi(t-\zeta y)}{(\zeta+1)^{\frac{k+3}{2}}}\)\right|_{\zeta=1}
\eea
For $k$ odd, we may apply the usual Leibniz rule to compute this multiple derivative, and get for an appropriately chosen coefficient $c_{\frac{k+1}{2}}$,
\bea
X(y,t)=\xi(t-y)+y\xi'(t-y)+...+c_{\frac{k+1}{2}}y^{\frac{k+1}{2}}\xi^{(\frac{k+1}{2})}(t-y)\label{linear}
\eea
For even $k$ we instead get an infinite series involving fractional derivatives,
\bea
X(y,t)=\sum_{p\leq \frac{k+1}{2}} c_p y^p \xi^{(p)}(t-y)
\eea
For both odd and even $k$ the coefficients satisfy the recursion relation
\bea
(p+1)(p-k-1)c_{p+1}=(2p-k-1)c_p\label{rec}
\eea

\section{Energy-momentum in the wave}
The Nambu action of a $k$-brane in $AdS_{d+1}$, 
\bea
S=-T\int dtd^{k-1}x\int_0^{\infty}dy \sqrt{-g}
\eea
is invariant under translations in the $d$ boundary spacetime directions $x^{\mu}$. The corresponding conserved energy-momentum charges are given by (in the same Monge gauge as in the previous section)
\bea
P_{\mu}=T\int_{t=x^0} d^{k-1}x\int_0^{\infty} dy\frac{1}{y^2}\eta_{\mu\nu}\sqrt{-g}g^{0A}\partial_{A}X^{\nu}
\eea
We compute these charges at a time $t$, but if these charges are time-independent then the precise choice of $t$ will not matter. These will be conserved charges provided no energy enters the membrane from the boundary string.
A non-BPS selfdual string will radiate energy-momentum until it has decayed into a BPS string. The mass of the BPS string is given through its central charge. If we assume that the string is BPS both before and after it has flucutated, it is natural to define the energy in the wave (on the membrane, whose boundary is the string), as the mass of the membrane minus its BPS mass. That means that the energy carried away by the wave is given by
\bea
P_0=T\int d^{k-1}x\(\int_0^{\infty} dy \frac{1}{y^{k+1}}\sqrt{-\hat{g}}\hat{g}^{00}-\int_0^{\infty} dy\frac{1}{y^{k+1}}\)
\eea
where we have defined $\hat{g}_{AB}\equiv y^2 g_{AB}$. The group $SO(2,d)$ is a symmetry of $AdS_{d+1}$. This symmetry is in the boundary theory realized as a conformal symmetry. If we fix a conformal gauge by defining the metric on the boundary as $\lim_{y\rightarrow 0} y^2 (ds)^2$, we get a residual symmetry which is the Lorentz group $SO(1,d-1)$. The energy-momentum vector therefore transforms as a vector under the Lorentz group of the boundary theory. It should therefore suffice to compute it in the galileian limit in the boundary theory. We propose that on the $AdS$ side the galileian limit corresponds to the linearized wave solution when it has propagated a far distance away from the boundary $y=0$. In this limit we get the energy in the membrane at $AdS$ time $t$ as
\bea
P_0=T\int d^{k-1}x \int_0^{\infty} dy y^{-(k+1)}\frac{1}{2}\((\partial_y X)^2+(\partial_t X)^2+\sum_{a=1}^{k-1}(\partial_a X)^2\)\label{te}
\eea
Writing the wave solution in terms of a Green's function as in Eq. (\ref{Green}), we get the energy as
\bea
P_0=T\int d^k x'\int d^k x'' G(x',x'')\xi(x')\xi(x'')
\eea 
where we have introduced\footnote{Taking the limit $t\rightarrow \infty$ can be justified if we assume that the string stops fluctuating after some time $t_1$. No more energy will radiate after that the string has stopped and therefore we may take $t\rightarrow \infty$ without getting any additional contribution to the total energy. The total energy radiated should then be independent of $t$ as long as $t>t_1$.}
\bea
G(x',x'')&\equiv&\lim_{t\rightarrow\infty}\int d^{k-1}x\int_0 dy \frac{1}{y^{k+1}}\Big(\partial_yK(y,x;x')\partial_yK(y,x;x'')\cr
&&+\partial_{t}K(y,x;x')\partial_{t}K(y,x;x'')+\sum_{a=1}^{k-1}\partial_{a}K(y,x;x')\partial^{a}K(y,x;x'')\Big)
\eea
We must be careful when changing the order of integrations. For $k=1$ we get
\bea
\int_0^{\infty} dy \int_{-\infty}^{t-y} dt'\int_{-\infty}^{t-y} dt'' = \int_{-\infty}^{t}dt'\int_{-\infty}^tdt''\int_{0}^{t-t'}dy\theta(t'-t'')+\Big(t'\leftrightarrow t''\Big)\label{Eq}
\eea
We get a triple-pole at $y=t-t'$ and another at $y=t-t''$. One of these poles is at the edge of the domain over which we integrate $y$, and the other is outside this domain and therefore gives no contribution. A triple-pole at the edge leads to a distribution $\partial^2\delta(y-(t-t'))$. If we would care about the normalizations, we would see that since we have a product of two Green's functions here, we also have a product of two `zero-constants' $c$. When we say that a triple-pole at the edge leads to a delta function distribution, we really use up one of these constants $c$ leaving us with just one constant $c$ left. The result we get is
\bea
G(t',t'')=c\frac{1}{|t'-t''|^{5}}\leadsto\partial^4\delta(t'-t'')
\eea
(To understand the last step, we take "$c\sim 1/\Gamma(-4)=0$".) This is also the only possibility by dimensional analysis and translational invariance.\footnote{The only other possibility $G(t',t'')=c\frac{1}{(t'-t'')^5}$ is ruled out since $G(x',x'')$ should be symmetric since it is `contracted' with $\xi(x')\xi(x'')$.} This yields the total energy radiated
\bea
P_0\sim\int_{-\infty}^t dt'\xi(t')\partial^4\xi(t')
\eea
Since we have assumed that $\xi(t)=0$, we may perform integrations by parts twice to arrive at the familiar Larmor formula for the radiated energy from a nonrelativistic particle \cite{Jackson},
\bea
P_0\sim\int_{-\infty}^t dt'\ddot\xi(t')^2.
\eea 
In this case we could have arrived at this result by inserting the explicit wave solution $X(y,t)=\xi(t-y)+y\xi'(t-y)$ into Eq. (\ref{te}), if we remember to stop the string at some time, $\xi(t')=0$ for $t'>t_1$ say, and then take $t>t_1$. The exact relativistic computation of this using the exact nonlinear wave solution in $AdS$ space was carried out in \cite{Mikhailov}, with the result 
\bea
P_{\mu}\sim \int d\xi_{\mu} \(D_{\tau}\xi_{\nu}\)^2
\eea
in covariant gauge $-\partial_{\tau}\xi_{\mu}\partial_{\tau}\xi^{\mu}=1$. Here $D_{\tau}=\partial_{\tau}$ denotes the covariant derivative on the particle world-line.
 
For $k=2$ we have a singularity at $y=\sqrt{(t-t')^2-(x-x')^2}$ of degree $4$, leading to $\partial^3\delta(y-\sqrt{(t-t')^2-(x-x')^2})$. In the limit $t\rightarrow \infty$ only one term survives (which is the one where $\partial_y^3$ hits the most singular factor in denominator), which yields
\bea
G(t',x';t'',x'')=c\frac{1}{|x'-x''|(t'-t'')^6}\leadsto \delta(x'-x'')\partial_{t'}^5\delta(t'-t'')
\eea
The total radiated energy is
\bea
P_0\sim\int dt' dx' \xi(t',x')\partial_{t'}^5\xi(t',x')\label{eq1}
\eea
For a rigid string (and more generally, for rigid objects in any dimension) we can obtain the radiated energy just by analytically continuing the exponent in the Larmor formula. We should thus rewrite the radiated energy as
\bea
P_0\sim \int dt'dx' \(\partial_{t'}^{5/2}\xi(t',x')\)^2\label{eq2}
\eea
and we expect this to be the time-integrated power that is radiated from the string. Both Eq. (\ref{eq1}) and Eq. (\ref{eq2}) yield the same {\sl{total}} energy though. These two expressions are not quite related by just a boundary term, but the equivalence of these two expressions becomes apparent if one goes to the Fourier space. (Recall that we may take $t\rightarrow \infty$.)

To the string is associated a central charge $Z_{\rho}$, which really is a Lorentz vector (it is also a $SO(5)$ R-symmetry vector, but that will not be important here). That is, given a string, we get a distinguished and conserved (time independent) direction which is given by the central charge. Parametrizing the string world-sheet by $\sigma^{\alpha}$ we may impose the covariant gauge $-\det(\partial_{\alpha}\xi_{\mu}\partial_{\beta}\xi^{\mu})=1$. We have a covariant derivative $D_{\alpha}$ on the string world-sheet. We may now construct a scalar that reduces to the space derivative to linear order in $\xi$ in Monge gauge,
\bea
E_x\equiv \frac{1}{\sqrt{Z_\rho Z^{\rho}\xi_{\kappa}\xi^{\kappa}}} Z_{\mu}\(D^{\alpha}\xi^{\mu}\)D_{\alpha}
\eea
The time derivative should covariantize to
\bea
E_t\equiv\sqrt{-D_{\alpha}D^{\alpha}-(E_x)^2}
\eea
Accordingly we propose the following covariant formula for the radiated energy-momentum
\bea
P_{\mu}\sim Z^{\rho}\int d\xi_{\mu}\wedge d\xi^{\rho} \((E_t)^{5/2}\xi_{\nu}(\sigma)\)^2
\eea
in covariant gauge. 

We expect this covariantized result to correspond to the nonlinear wave solution. For a rigid straight object with $k$ odd, the nonlinear wave should be given by
\bea
X^{\mu}&=&\xi^{\mu}(\tau)+y \dot{\xi}^{\mu}(\tau)+...+c_{\frac{k+1}{2}}y^{\frac{k+1}{2}}\partial^{\frac{k+1}{2}}\xi^{\mu}(\tau)
\eea
in the gauge where $-\dot{\xi}_{\mu}(\tau)\dot{\xi}^{\mu}(\tau)=1$, since this reduces to the nonlinear wave obtained in \cite{Mikhailov} if we take $k=1$. Furthermore, in the Monge gauge $X^0=t$, $\xi^0=t'$, the nonlinear wave above reads
\bea
t&=&t'+\gamma y+...\cr
X^i&=&\xi^i(t')+\gamma y \dot{\xi}^i(t')+...
\eea
where $\gamma\equiv \(1-\dot{\xi}(t')^2\)^{-1/2}$ is the usual relativistic gamma factor. We deduce that in the galileian limit this reduces to the linear wave solution Eq. (\ref{linear}). There should be a corresponding nonlinear wave for rigid straight objects in even dimensions $k$, though the boundary condition can not be seen in the series expansion where we expand in odd powers of $y$. But it may be seen in an integral representation of the wave. We leave it as an open question to find the nonlinear wave corresponding to generic boundary conditions.

\section{Radiation in classical electrodynamics}
A string which is charged under a $U(1)$ gauge group may couple to a two form gauge potential $B$ via $\int B$ where the integration is over the string world-sheet. Denoting the field stength by $H=dB$, the total action for this system is
\bea
S=-\frac{1}{2}\int H\wedge *H+\int B-T\int d^2\sigma \sqrt{-g}
\eea
where we parametrize the string world-sheet by $\sigma^{\alpha}$ which leads to the  induced metric tensor $g_{\alpha\beta}$. If this action were to model a selfdual string, then (apart from the usual redefinition of the field strength to include the magnetic charge in the Bianchi identity) the string tension $T$ would be set by the norm of the central charge associated with the string, which in turn is determined by the scalar fields in the tensormultiplet at infinity. If in addition we have external tensormuliplet fields, we get the string tension as the sum of two contributions. The first contribution is from the self-fields that the string produces itself, which is to be evaluated at infinity. This will be the non-dynamical part of the string tension. If we assume that the vacuum expectation value of these scalar self-fields points in the internal direction $\hat{n}^a$ (where $a$ is a vector index under the $SO(5)$ R-symmetry), then the second contribution to the string tension will be given by the sum of all the external scalar fields projected onto that direction $\hat{n}^a$ when evaluated at the location of the string under consideration.\footnote{This relies on the assumption that the scalar self-field the string produces itself points in the $\hat{n}$-direction also at the string location. If this assumption would eventually turn out to be false, then any external scalar field component could interact with the string.} Such external fields could come from other strings, or from tensormultiplet waves coming against the string from infinity. The latter contribution to the string tension will thus in general be spacetime dependent. The string tension is thus dynamical insofar as the external tensorfield are treated as dynamical fields. 

But here we will not assume a selfdual string, nor a dynamical string tension. Our result for the radiated energy will not depend crucially on these assumptions. The radiation from the field coming from the magnetic charge of the selfdual string will yield the same contribution to the radiated energy as the electric charge, and so will the scalar field that radiates from the string. We will therefore only consider the electromagnetic field radiated from an electrically charged string, which means that we can adopt the familiar results in classical electrodynamics (see for instance \cite{Jackson}).

The physical stress tensor may be obtained as the variation of this action with respect to a spacetime dependent six-dimensional metric tensor $G_{\mu\nu}(x)$. We will work in flat Minskowski spacetime, so after the variation we put $G_{\mu\nu}=\eta_{\mu\nu}$. We then get the total stress tensor as 
\bea
T^{\mu\nu}&=&T_{e.m.}^{\mu\nu}+T_{string}^{\mu\nu}
\eea
with
\bea
T_{e.m.}^{\mu\nu}&=&H^{\mu\kappa\tau}H^{\nu}{}_{\kappa\tau}-\frac{1}{6}\eta^{\mu\nu}H_{\kappa\tau\rho}H^{\kappa\tau\rho}\cr
T_{string}^{\mu\nu}&=&-\frac{T}{2}\int d^2\sigma \sqrt{-g} \delta^6(x-X(\sigma)) g^{\alpha\beta}\partial_{\alpha} X^{\mu}\partial_{\beta} X^{\nu}
\eea
The total stress tensor is conserved, $\partial_{\mu}T^{\mu\nu}=0$, when using the equations of motion for the electromagnetic field and the string. But the stress tensors associated with the field alone is not conserved,
\bea
\partial_{\mu}T^{\mu\nu}_{e.m.}=-\partial_{\mu}T^{\mu\nu}_{string}=\int dX_{\kappa}\wedge dX_{\rho}H^{\nu\kappa\rho}(X)
\eea
This is simply reflects the fact that energy radiates from the string into the field.\footnote{To derive this non-conservation of the field stress tensor we used the Maxwell equations of motion and the Bianchi identity, and for the string stress tensor we used the string equation of motion.}

The rate of energy radiated from the string in unit time is now obtained by considering the volume $V={\bf{R}}^5-W$ at a fixed time, where $W$ is a tubular neighborhood of the string. In this region we have no sources so the field stress tensor is conserved, and hence upon integrating over $V$
\bea
\frac{dP_0}{dt}=\int_{\partial W} dS_i T^{i}{}_0
\eea
where $dS_i$ is the directed area element on the boundary of $W$ and $P_0=\int_V d^5 x T^0{}_0$ is the energy that has radiated out from the tubular neighborhood of the string. This is known as the Poynting theorem, and 
\bea
T^{0i}=H^{0\mu\nu}H^{i}{}_{\mu\nu}
\eea
is the Poynting vector.

Our aim now is to compute this Poynting vector and radiated energy. We must then first solve the Maxwell equations for the electromagnetic field, at least in the far field region, given a prescribed motion of the string. Making the ansatz
\bea
B_{\mu\nu}(x)=\int d^2\sigma \(\dot{X}_{\mu}X'_{\nu}-\dot{X}^{\nu}X'_{\mu}\)J(x-X(\sigma))
\eea
and imposing the gauge fixing condition $\partial^{\mu}B_{\mu\nu}=0$, we find that this solves the Maxwell equations provided
\bea
\partial^{\mu}\partial_{\mu}J(x)=\delta^6(x)
\eea
The retarded solution is 
\bea
J(x^0,{\bf{x}})=-\frac{4}{3\Omega_4}\Theta(x^0)\delta'\(x^2\)
\eea
Here $x^2\equiv\eta_{\mu\nu}x^{\mu}x^{\nu}$, and this will be a Lorentz scalar despite the theta function due to the delta function which constrains $x$ to lie on the light-cone. The Lienard-Wiechert potential is now
\bea
B^{\mu\nu}(x)=\int dX^{\mu}\wedge dX^{\nu}\delta'\((x-X)^2\)
\eea
Here we can use
\bea
\partial^{\mu}\delta\((x-X(\sigma))^2\)&=&2{(x-X(\sigma))^{\mu}}\delta'((x-X(\sigma))^2),\cr
\delta'\((x-X(\sigma))^2\)&=&\frac{1}{2\dot{X}(\sigma).(x-X(\sigma))}\frac{\partial}{\partial \tau}\delta\((x-X(\sigma))^2\)
\eea
to get, after some integrations by parts, 
\bea
B^{\mu\nu}(x)=\int d\sigma \frac{1}{2\dot{X}.(x-X)}\frac{\partial}{\partial\tau}\(\frac{\dot{X}^{\mu}X'^{\nu}}{2\dot{X}.(x-X)}\)
\eea
and
\bea
\partial^{\rho}B^{\mu\nu}(x)=\int d\sigma \frac{1}{\dot{X}.(x-X)}\frac{\partial}{\partial\tau}\(\frac{1}{\dot{X}.(x-X)}\frac{\partial}{\partial\tau}\(\frac{\dot{X}^{\mu}X'^{\nu}(x-X)^{\rho}}{\dot{X}.(x-X)}\)\)
\eea
These expressions are to be evaluated at the retarded time $\tau=\tau_0$ specified by the conditions $(x-X(\sigma))^2=0$, $x^0-X^0(\sigma)>0$.

\subsection{Electromagnetic radiation in the far field region}
We now impose the Monge gauge $X^0=t$, $X^5=x$, $X^{i=1,2,3,4}=\xi^i(t,x)$. Using the above formula for the retarded field strength, we get the following (linearized) solution in the radiation zone $R\equiv \sqrt{x^i x^i}>>\xi$,
\bea
H^{ijk}(x)&=&O(\xi^2)\cr
H^{0ij}(x)&=&\int dx \frac{\ddot{\xi'}^ix^j-\ddot{\xi'}^jx^i}{(x^0-t)^3}+...\cr
H^{ij5}(x)&=&\int dx \frac{x^i\dddot{\xi}^j-x^j\dddot{\xi}^i}{(x^0-t)^3}+...\cr
H^{0j5}(x)&=&-2\int dx\frac{x^j}{(x^0-t)^4}\cr
&&+\int dx\(\frac{\dddot{\xi}^j}{(x^0-t)^2}+\frac{(x^5-x)\ddot{\xi'}^j}{(x^0-t)^3}-\frac{x^jx.\dddot{\xi}}{(x^0-t)^4}\)+...
\eea
Here $t=x^0-\sqrt{(x^5-x)^2+R^2}$ is the retarded time. The dots represent terms that are either of higher order in $1/R$ (and therefore negligible in the radiation zone) or $O(\xi^2)$.  Fourier decomposing 
\bea
\xi(t,x)=\int dEdk \xi(E,k)e^{i(Et+kx)}
\eea
we will get integrals of the form
\bea
\int dx \frac{e^{if(x)}}{\(R^2+(x-x^5)^2\)^p}
\eea
with exponent 
\bea
f(x)=-E\sqrt{(x^5-x)^2+R^2}+kx
\eea
Changing the variable $x$ to a dimensionless variable $s=\frac{x-x^5}{R}$ and defining an angle $\theta$ as $k=E\cos\theta$ it becomes
\bea
f(s)=kx^5+R\(-E\sqrt{1+s^2}+ks\)
\eea
Since we are interested only in the far-distance (large $R$) behavior, we can compute these integrals using steepest descent. That means evaluating the integrand at the extremum point of the exponent $f(s)$. There is just one extremum point, which is at $s=s_0=1/\tan\theta$. We then Taylor expand the exponent about the extremum point and keep only the quadratic term in this expansion to get a gaussian integral. The result is
\bea
H^{ij5}(x)&=&\int dEdke^{iE(x^0+x^5\cos\theta-R\sin\theta)} \frac{(iE)^{5/2}}{R^{3/2}}\frac{x^i\xi^j-x^j\xi^i}{R}\sin^{9/2}\theta \cr
H^{0j5}(x)&=&\int dEdke^{iE(x^0+x^5\cos\theta-R\sin\theta)}
\frac{(iE)^{5/2}}{R^{3/2}}\(\xi^j-\frac{x^j x.\xi}{R^2}\)\sin^{11/2}\theta
\eea
Rather similar results were obtained in \cite{Henningson} for the radiated fields from a bosonic selfdual string.

\subsection{Energy radiated}
The radiated energy-momentum is given by
\bea
P_{\mu}=\int dx^0 dx^5 \int d\Omega R^2 x^k T^{k}{}_{\mu}(x)
\eea
where $d\Omega$ is the volume element of the unit three sphere, and we may take the radius of the three sphere surrounding the string to be arbitrarily large, $R>>\xi$ since no energy that radiates from the string can disappear on its way out to infinity. For the radiated energy we need 
\bea
T^{i}{}_{0}(x)&=&H_{0j5}(x)H^{ij5}(x)+H_{0ik}H^{ijk}(x)
\eea
The second term here is in $O(\xi^3)$ so we will neglect it. In principle we could get a contribution in $O(\xi^2)$ by multiplying $O(\xi^2)$ terms in $H^{ij5}$ with the $O(1)$ term (corresponding to a static straight string) $H_{0j5}\sim x_j/R^4$. But doing this gives just the contribution $\sim x^i x^j H^{ij5}(x) =0$ to the energy since $H^{ij5}$ is antisymmetric. 
When inserting the linearized solution for the field strength into the expression for the total radiated energy $P_0$ and replace $\cos\theta$ with $k/E$, we find that the integral over $x^0$ produces $\delta(E+E')$, and the integral over $x^5$ produces $\delta(k+k')$. This yields the following result for the radiated energy
\bea
P_0=\int dE dk E^5\(1-k^2/E^2\)^{5} \xi^i(E,k).\xi^i(-E,-k)+O(\xi^3)
\eea
The radiated momentum is $O(\xi^3)$. In position space we may write this as 
\bea
P_0=\int dt'dx'\(\partial^{5/2}_{t'}\(1-\partial_{x'}^2\partial_{t'}^{-2}\)^{5/2}\xi(t',x')\)^2
\eea
which is to be compared with Eq. (\ref{eq2}).

\vskip 0.5truecm
{\sl{Acknowledgements}}: 
I have benefited from discussions with Kjell Holm\aa ker.

\end{document}